\documentclass[twocolumn]{aastex631}
\hypersetup{colorlinks,linkcolor={cyan},citecolor={cyan},urlcolor={cyan}} 

\usepackage{lineno}

\usepackage{soul}

\usepackage{color}

\def\kms{\textit{$\rm km\ s^{-1}$}}

\def\ergs{\textit{$\rm erg\ s^{-1}$}}

\def\lbol{\textit{$L_{\rm bol}$}}

\def\ss{\textit{$\rm S\acute{e}rsic$}}
\def\thetae{\textit{$\theta_E$}}
\def\chiv{\textit{$\chi_\nu^2$}}
\def\q{\textit{$q$}}

\def\ebvline{\textit{$E(B-V)_{\rm line}$}}

\def\n{\textit{$n$}}
\def\mbh{\textit{$M_{\rm BH}$}}
\def\msun{\textit{$M_\odot$}}
\def\m{\textit{$M_{\star}$}}
\def\logm{\textit{${\rm log}\,M_{\star}$}}

\def\ha{\textit{${\rm H}\alpha$}}
\def\re{\textit{$R_{\rm e}$}}

\def\thetae{\textit{$\theta_{E}$}}
\def\gm{\textit{$\gamma$}}

\begin{document}

\title{Varstrometry for Off-nucleus and Dual sub-Kpc AGN (VODKA). SDSS J1608+2716:  A Sub-arcsec Quadruply Lensed Quasar at $z=2.575$}

\author[0000-0002-1605-915X]{Junyao Li}
\affiliation{Department of Astronomy, University of Illinois at Urbana-Champaign, Urbana, IL 61801, USA}
\correspondingauthor{Junyao Li}
\email{junyaoli@illinois.edu}

\author[0000-0003-0049-5210]{Xin Liu}
\affiliation{Department of Astronomy, University of Illinois at Urbana-Champaign, Urbana, IL 61801, USA}
\affiliation{National Center for Supercomputing Applications, University of Illinois at Urbana-Champaign, Urbana, IL 61801, USA}

\author[0000-0003-1659-7035]{Yue Shen}
\affiliation{Department of Astronomy, University of Illinois at Urbana-Champaign, Urbana, IL 61801, USA}
\affiliation{National Center for Supercomputing Applications, University of Illinois at Urbana-Champaign, Urbana, IL 61801, USA}

\author[0000-0003-3484-399X]{Masamune Oguri}
\affiliation{Center for Frontier Science, Chiba University, Chiba 263-8522, Japan}
\affiliation{Department of Physics, Graduate School of Science, Chiba University, Chiba 263-8522, Japan}

\author[0000-0001-7681-9213]{Arran C. Gross}
\affiliation{Department of Astronomy, University of Illinois at Urbana-Champaign, Urbana, IL 61801, USA}

\author[0000-0001-6100-6869]{Nadia L. Zakamska}
\affiliation{Department of Physics and Astronomy, Johns Hopkins University, Baltimore, MD 21218, USA}

\author[0000-0002-9932-1298]{Yu-Ching Chen}
\affiliation{Department of Astronomy, University of Illinois at Urbana-Champaign, Urbana, IL 61801, USA}
\affiliation{National Center for Supercomputing Applications, University of Illinois at Urbana-Champaign, Urbana, IL 61801, USA}

\author[0000-0003-4250-4437]{Hsiang-Chih Hwang}
\affiliation{School of Natural Sciences, Institute for Advanced Study, Princeton, NJ 08540, USA}

\begin{abstract}
We report {\it{Hubble Space Telescope}} (HST) Wide Field Camera 3 (WFC3) deep IR (F160W) imaging of SDSS J1608+2716. This system, located at a redshift of $z=2.575$, was recently reported as a triple quasar candidate with subarcsecond separations ($\sim0.25''$) based on selection from Gaia astrometry and follow-up Keck adaptive optics-assisted integral field unit spectroscopy. Our new HST deep IR imaging reveals the presence of a fourth point-like component located $\sim0.9''$ away from the triple system. Additionally, we detect an edge-on disk galaxy located in between the four point sources. The entire system exhibits a characteristic cusp structure in the context of strong gravitational lensing, and the observed image configuration can be successfully reproduced using a lens model based on a singular isothermal ellipsoid mass profile. These findings indicate that this system is a quadruply lensed quasar.
Our results highlight the challenges associated with identifying dual/multiple quasars on $\sim$kpc scales at high redshifts, and emphasize the crucial role of deep, high-resolution IR imaging in robustly confirming such systems. 
\end{abstract}

\section{Introduction}
In the context of hierarchical structure formation of the universe, galaxy mergers play a pivotal role in shaping the evolutionary pathways of star formation and structural properties of galaxies, and in driving the gas inflows toward the galaxy center to concurrently fuel the growth of their central supermassive black holes (SMBHs). The existence of multiple SMBHs in a galaxy merging system is expected to be common since almost all massive galaxies harbor a central SMBH in the local universe \citep[e.g.,][]{Kormendy2013}. Dual/multiple active galactic nuclei (AGNs) are a rare population in which more than one SMBHs are actively accreting matter and emitting copious amounts of energy simultaneously (dual fraction $\sim10^{-3}-10^{-4}$ among all quasars, i.e., $L_{\rm bol}\gtrsim 10^{45}\,{\rm erg\,s^{-1}}$, at $1\lesssim z \lesssim 3$; e.g., \citealt{Silverman2020, Shen2023}). These dual/multiple AGNs, which eventually culminate in the gravitational wave emissions of in-spiralling SMBHs, offer a unique window to test the theory of dynamical evolution of galaxy and SMBH mergers, and to study the processes governing galaxy transitions (e.g., feedback from multiple AGNs) and the still elusive role of mergers in fueling SMBH growth. 

The identification and study of close ($\sim$kpc) separation dual AGNs have been a significant scientific pursuit in recent years \citep[e.g.,][]{Liu2010, Silverman2020, Shen2021, Shen2023, Tang2021, Mannucci2022, Chen2022hst, Chen2023, Gross2023a}. How the abundance of them changes as a function of separation and luminosity provides critical constraints on the theoretical models of SMBH pair evolution \citep[e.g.,][]{Shen2023}. Of particular interest is searching for dual AGNs at cosmic noon (i.e., $1\lesssim z \lesssim 3$), the primary epoch of massive galaxy and SMBH formation where galaxy mergers are more frequent \cite[e.g.,][]{Duncan2019}. However, such systems are difficult to find given the stringent resolution requirement (e.g., subarcseconds for kpc scale separations). In recent years, two novel techniques based on the Gaia satellite have been proposed to break the resolution limit and efficiently discover kpc-scale dual quasars beyond $z\gtrsim1$ by leveraging Gaia's excellent point spread function (PSF) and superb astrometry precision ($\sim1$ mas). The varstrometry technique capitalizes on the ubiquitous stochastic variability of AGNs and identifies light centroid jitter caused by asynchronous variability from unresolved AGN pairs \citep{Hwang2020, Shen2021}, while the Gaia Multi Peak (GMP) method searches for multiple peaks in the Gaia light profiles of  unresolved AGNs \citep{Mannucci2022}. These approaches have proved to be very efficient in selecting multiple point-like sources with subarcsecond separations \citep[e.g.,][]{Chen2022hst}. The critical next step is to confirm their physical nature.

Multiple sources with close separations could be genuine quasar pairs/multiples, star-quasar superpositions, or gravitationally lensed single quasars which are of interest for many cosmological applications \citep{Treu2010}. Comprehensive multiwavelength follow-up observations are usually required to distinguish one from the others \citep[e.g.,][]{Liu2019, Chen2023, Gross2023}. Despite the continued observational effort {of finding promising candidate kpc-scale dual AGNs at $z>1$ \citep[e.g.,][]{Yue2021, Glikman2023}}, only a few systems with separations below 5~kpc are robustly confirmed \citep{Junkkarinen2001, Mannucci2022, Chen2023}. 

In this paper, we focus on SDSS J160829.23+271626.7 (hereafter SDSS J1608+2716 for short), which is a spectroscopically confirmed Type 1 (i.e., broad-line) quasar at $z=2.575$ \citep{Dawson2013}. It has been recently reported as a close-separation triple quasar candidate based on adaptive optics (AO)-assisted Keck integral field spectroscopy (IFU; \citealt{Ciurlo2023}), originally selected as a GMP source \citep{Mannucci2022}. The Keck observations unambiguously revealed the presence of three distinct components with separations of $\sim0.25''$ ($\sim2$ kpc), each emitting a broad \ha~line with FWHM of $\sim5000\,\kms$. Notably, two components display similar line profiles while the third component exhibits slight differences in both line width and centroid. The non-detection of a foreground lens in the AO data makes it a promising candidate for a genuine triple quasar system, although the possibility of lensing can not be ruled out based on the shallow Keck imaging \citep{Ciurlo2023}.

Here we report HST/WFC3 IR imaging observation of J1608. Our data reveals the presence of a fourth point-like component located $\sim0.9''$ from the triple, alongside an edge-on disk galaxy positioned between the four point sources. We show that based on the currently available data, the most plausible interpretation for this system is a single quasar being quadruply lensed into a cusp configuration (i.e., sources near a cusp of the caustic curve produce a configuration where three of the images lying close together on one side of the lens galaxy; \citealt{Keeton2003_cusp}) by a foreground disk galaxy. Throughout this paper we adopt a flat $\Lambda$CDM cosmology with $\Omega_{\Lambda}=0.7$ and $H_{0}=70\,{\rm km\,s^{-1}Mpc^{-1}}$. Magnitudes are given in the AB system.

\section{Observations}
We observed SDSS J1608+2716 (J2000 coordinates: RA = 16:08:29.23, DEC = +27:16:26.74) with HST/WFC3 in the IR F160W band (i.e., $H$ band, with central wavelength $\lambda = 15436\rm\ \AA$ and effective width of $2750\rm\ \AA$) as part of the VODKA program in Cycle~30 on 2023 April 23 (Program ID: GO-17287; PI: X. Liu) with a four-point dither pattern. The integrated exposure time was 2062 s. The individual exposures were dithered, cosmic ray and hot pixel rejected, and combined with {\tt{DrizzlePac}} \citep{Hoffmann2021} with a {\tt{pixfrac}} of 0.8. The final combined image has an output pixel scale of $0.065''$. 

Since no field star was available in the field of view (FOV) of J1608 for PSF construction, we searched the archival data for stars that used the same dither pattern and observing mode as ours to build the PSF. The only two bright stars observed in close date to J1608 originated from our VODKA programs GO-17287 and GO-17269 (PI: X. Liu). These stars, nominated as star0841 and star2122, were observed on 2023 March 28 and April 12 in the frame of SDSS J0841+4825 and SDSS J2122-0026, respectively. We constructed the effective PSF model (ePSF; \citealt{Anderson2000}) using the two stars through the {\tt{EPSFBuilder}} method available in the python package {\tt{photutils}} \citep{Bradley2022}. In order to assess the impact of PSF mismatch on our results, we further included five ePSF models built from stars observed in our Cycle 29 program GO-16892 (PI: X. Liu). Each of these additional PSF models was constructed from at least two stars in the FOV of five SDSS quasar targets.
The PSF image was drizzled to the same pixel scale as J1608. Each image was background subtracted using the {\tt{SExtractorBackground}} algorithm available in {\tt photutils}.

\begin{figure*}
\includegraphics[width=\linewidth]{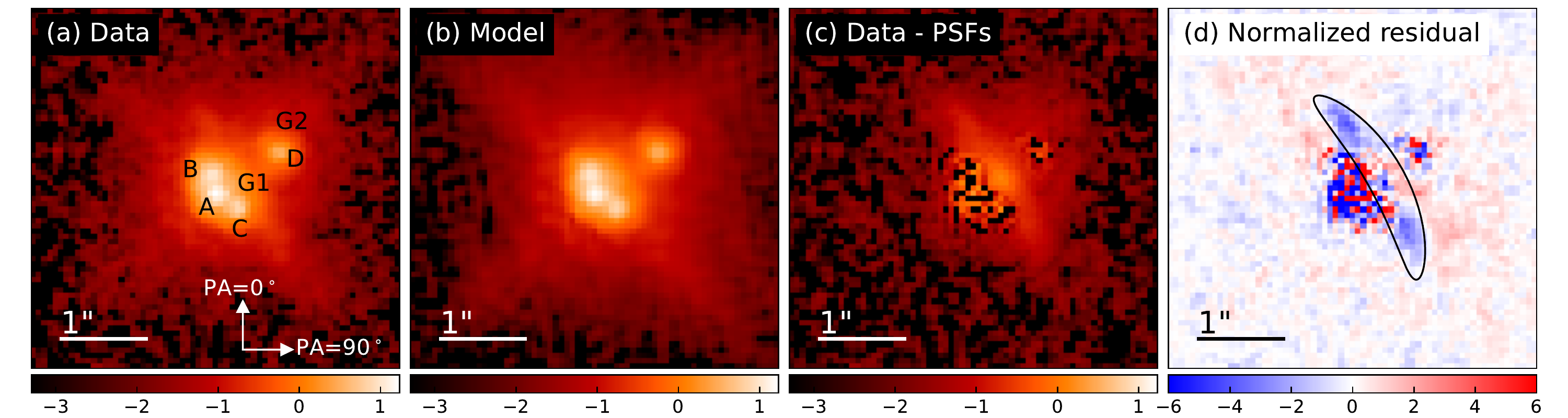}
\caption{Quasar-galaxy decomposition result using a 4 PSF (for ABCD) + 2 \ss~(for G1 and G2) model: (a) observed image with a logarithm scaling; (b) best-fit model image convolved with the PSF; (c) data minus the 4 PSFs (i.e., the pure galaxy image); (d) fitting residuals (model - data) divided by the error map. The black curve marks the warped disk of G1.}
\label{fig:decomp}
\end{figure*}

\section{Image Analysis}
The HST IR image of J1608 (Figure \ref{fig:decomp}) exhibits four bright point-like components (labeled as A, B, C, and D) and two extended components (G1 and G2). One extended component G1 appears to be an edge-on disk-like galaxy located between ABCD, while G2 exhibits extended emissions in the outskirts but is outshined by the point source D in the central region. The Keck AO-assisted IFU spectrum in \cite{Ciurlo2023}, which had a FOV of $1.6''\times3.2''$ that covers D and G2, did not report any signals detected at their positions. Nor was a foreground lens detected in \cite{Ciurlo2023}, leading the authors to conclude that ABC is most likely to be a bona-fide triple quasar system. Our deep HST observation with a larger FOV shows that the image configuration closely resembles a classical cusp structure for a quadruply lensed quasar, with the putative foreground lens clearly detected between the four source images. This system configuration makes the lensing scenario the most likely interpretation for this multiple system. In the following sections, we conduct detailed image modeling to better constrain the physical nature of these components and explore potential interpretations for the observed image configuration.

\subsection{Quasar-galaxy decomposition}
\label{subsec:decomp}
We perform two-dimensional (2D) surface brightness decomposition with {\tt lenstronomy}\footnote{\url{https://github.com/lenstronomy/lenstronomy.}} \citep{Birrer2018, Birrer2021} to disentangle several blended components. {\tt Lenstronomy} is a multipurpose package initially designed for strong gravitational lensing analysis through forward modeling using a particle swarm optimization algorithm. Its high-level of flexibility enables us to deactivate the lensing module and decompose images into quasar and galaxy components based on 2D profile fitting. 

Our baseline model includes four PSF models to represent the four point source components (i.e., ABCD), and two \ss~profiles convolved with the PSF to fit the extended galaxies G1 and G2. The 1D projection of the \ss~profile is parameterized as 
\begin{equation}
I(r) = I_e\, {\rm exp}\left(-b_n \left[\left(\frac{R}{\re}\right)^{1/n} - 1\right]\right),
\end{equation}
where \re~is the effective radius along the major axis, \n~is the \ss~index with the constant $b_n$ being uniquely determined for a given \n, and $I_e$  represents the flux intensity at \re. Two additional parameters, the minor-to-major axis ratio \q~and the position angle PA (as defined in Figure \ref{fig:decomp}), are included to describe the shape and orientation of the galaxies. 

\begin{table*}
\renewcommand{\arraystretch}{1.2}
\caption{Quasar-galaxy decompositoin result using 4 PSFs for ABCD and 2 \ss~models for G1 and G2. The positions of each component ($\Delta$X and $\Delta$Y) are given in relative units to the image center. }
\centering
\begin{tabular}{ccccccc}
\hline
\hline
Params & A & B & C & D & G1 & G2\\
\hline
mag & $19.93_{-0.00}^{+0.04}$ & $20.33_{-0.03}^{+0.02}$ & $20.77_{-0.00}^{+0.03}$ & $21.58_{-0.01}^{+0.05}$ & $20.78_{-0.04}^{+0.11}$ & $22.89_{-0.13}^{+0.07}$\\
$\rm \Delta X \,(^{\prime\prime})$ & $0.01_{-0.00}^{+0.00}$ & $-0.04_{-0.00}^{+0.00}$ & $0.25_{-0.00}^{+0.00}$ & $0.75_{-0.00}^{+0.00}$ & $0.30_{-0.00}^{+0.01}$ & $0.73_{-0.01}^{+0.01}$\\
$\rm \Delta Y \,(^{\prime\prime})$ & $-0.07_{-0.00}^{+0.00}$ & $0.17_{-0.00}^{+0.00}$ & $-0.23_{-0.00}^{+0.00}$ & $0.43_{-0.00}^{+0.00}$ & $0.12_{-0.00}^{+0.01}$ & $0.47_{-0.01}^{+0.03}$\\
$R\,(^{\prime\prime})$ & ... & ... & ... & ...& $0.40_{-0.05}^{+0.02}$ & $0.30_{-0.02}^{+0.06}$\\
$n$ & ... & ... & ... & ...& $4.03_{-0.05}^{+0.39}$ & $1.98_{-0.16}^{+0.42}$\\
$q$ & ... & ... & ... & ...& $0.30_{-0.00}^{+0.03}$ & $0.82_{-0.04}^{+0.01}$\\
PA & ... & ... & ... & ...& $-57.21_{-0.01}^{+0.46}$ & $22.12_{-70.52}^{+5.70}$\\
\hline\\
\end{tabular}
\label{table:decomp}
\end{table*}

The decomposition results, obtained using the ePSF model constructed from star0841 and star2122, are presented in Figure \ref{fig:decomp}. The fitted parameters and their associated uncertainties are derived from the 16th, 50th, and 84th percentiles of the best-fit parameters obtained from fitting with different PSF models, as summarized in Table \ref{table:decomp}. The consistency of the best-fit values from different runs (i.e., small parameter uncertainties) demonstrates that PSF mismatch does not impact our results.
The decomposed PSF magnitudes for components A, B, C, and D are $19.93_{-0.00}^{+0.04}$, $20.33_{-0.03}^{+0.02}$, $20.77_{-0.00}^{+0.03}$, and $21.58_{-0.01}^{+0.05}$ mags, respectively. By adopting the total bolometric luminosity  ($\lbol\sim10^{46.2}\,\ergs$) estimated for this system within the SDSS $3''$-diameter fiber \citep{Wu2022}, the individual \lbol~of ABCD are approximately $1-7 \times 10^{45}\,\ergs$ based on the flux ratio obtained from image decomposition. The non-detection of D in \ha~could be due to its relative faintness, with expected \ha~flux being only half of that observed for C (i.e., the faintest \ha~emitting source detected in \citealt{Ciurlo2023}) based on our decomposition result, thus comparable to the noise level in \cite{Ciurlo2023}. 

We also consider the possibility that component D is not an AGN but rather a compact bulge-dominated galaxy with an extended disk. To test this hypothesis, we replace the PSF+\ss~profiles for D+G2 in our baseline model with two \ss~profiles (disk+bulge, convolved with the PSF) and assess the robustness of the point-source detection via comparison of the goodness-of-fit parameter, specifically the reduced chi-squared value \chiv. In all the runs using different PSF models, the \chiv~value of the \ss+\ss~model is worse than the baseline fit even with more free parameters. Moreover, the \ss+\ss~fit fails to converge to physically realistic parameters (\re~for the ``bulge'' hits the 1-pixel lower limit). Therefore, we conclude that component D is a point source, although spectroscopy is required to robustly confirm its nature.

After subtracting the bright point sources from the image (Figure \ref{fig:decomp}), the underlying galaxy component exhibits a complex morphology. In general, the system consists of an edge-on disk-like galaxy G1 with a central bulge as described by $q\sim0.3$ and $n\sim4.0$. Its total magnitude is $\sim20.8$ mag from our fitting, which is $\sim3$ times fainter in surface brightness than C, placing it just below the Keck detection limit in \cite{Ciurlo2023}. The disk of G1 exhibits a classic U-type warped morphology, which cannot be appropriately described by a smooth \ss~profile, as indicated by the black curve in Figure \ref{fig:decomp}. This feature suggests that G1 is likely being tidally stripped by its satellite galaxies \citep[e.g.,][]{Reshetnikov2002}.
Moreover, G1 is surrounded by extended emissions that likely originated from additional galaxies. The presence of such a complex galaxy morphology suggests two possible scenarios to explain the observed image configuration: 1) ABCD is a single quasar being quadruply imaged via strong lensing, and the extended emissions around G1 are from the lensed quasar host galaxy; and 2) it is a system of four distinct/individual quasars participating in a merger where the additional galaxies are their multiple host galaxies at the same redshift. Notably, the triple ABC are offset from the brightest galaxy G1 and instead appears to be centered on the outer extended emissions. In the following sections, we delve into these possibilities in detail.

\subsection{Lensing scenario}
\label{subsec:lens}

\begin{table*}
\centering
\caption{Lens modeling result of the mass profile of the deflector and the magnification ($\mathcal{M}$) at the position of the point sources}. Fixed parameters are labeled by *.
\begin{tabular}{ccccccccccc}
\hline
\hline
Model & $\thetae''$ & $\gm$ & $q$ & PA$^\circ$ & $\rm \Delta\,X''$ & $\rm \Delta\,Y''$ & $\mathcal{M}_{\rm A}$ & $\mathcal{M}_{\rm B}$ & $\mathcal{M}_{\rm C}$ & $\mathcal{M}_{\rm D}$\\
\hline
G1 & $0.42_{-0.00}^{+0.00}$ & 2.0* &
$0.62_{-0.01}^{+0.01}$ & 
$-54.82_{-0.02}^{+0.08}$ &
$0.31_{-0.00}^{+0.00}$ &
$0.11_{-0.00}^{+0.00}$ &
$15.05_{-0.54}^{+1.35}$ &
$-7.97_{-0.65}^{+0.36}$ &
$-6.85_{-0.70}^{+0.30}$ &
$2.57_{-0.05}^{+0.14}$\\
\hline\\
\end{tabular}
\label{table:lens}
\end{table*}

We fit the image configuration with {\tt lenstronomy} to examine the lensing scenario. The lens model is constrained by the positions of the lensed quasar as well as the spatially extended surface brightness distribution of its host galaxy. The flux ratios of the quasar are not employed as constraints in the modeling process as they can be influenced by microlensing effects and differential dust extinction \citep[e.g.,][]{Keeton2006}. Given the presence of multiple degeneracies in the lens modeling, it is unlikely that our single-band imaging can provide a unique solution of the mass distribution \citep[e.g.,][]{Schneider2013}. Therefore, the primary objective of this study is to examine whether the observed image configuration can be reproduced by a lens model with a minimum number of parameters, instead of achieving a precise fit to constrain the mass profile and magnification of the deflectors. To accomplish this goal, we impose several physically-motivated priors on the choice of mass and light profiles and parameter ranges.

The combined (e.g., disk + bulge + dark matter) mass profile of the deflector G1 is fitted by a singular isothermal
ellipsoid (SIE) profile. The SIE model has a radial mass density profile of $\rho\propto r^{-\gamma}$ where the power-law slope $\gamma$ is fixed to 2.0 \citep{Auger2010}. The dimensionless projected surface mass density (i.e., convergence $\kappa$) is given by 
\begin{equation}
    \kappa (x, y) = \frac{3-\gamma}{2} \left( \frac{\thetae}{\sqrt{qx^2+y^2/q}}\right)^{\gm-1},
\end{equation}
where \thetae~is the (circularized) Einstein radius and $q$ is the minor/major axis ratio. 
The adopted light profiles for G1 and ABCD are the same as in Section \ref{subsec:decomp}. The host galaxy of the unlensed quasar is modeled using a \ss~profile with the same centroid as the quasar point source. This time we refrain from adding an additional \ss~profile for G2 to examine if the extended emissions surround G1 can be explained by the lensed quasar host galaxy. Motivated by \cite{Ertl2023}, we adopt a Gaussian prior for the centroid of the mass profile of G1 based on the centroid of its corresponding light profile with a standard deviation of $0.065''$ (i.e., 1 pixel). The PA of G1's mass profile is allowed to vary within $10^\circ$ to that of its light profile \citep{Ertl2023} and its axis ratio cannot be smaller than that of the light profile by more than 0.1 \citep{Schmidt2023}.

\begin{figure*}
\centering
\includegraphics[width=\linewidth]{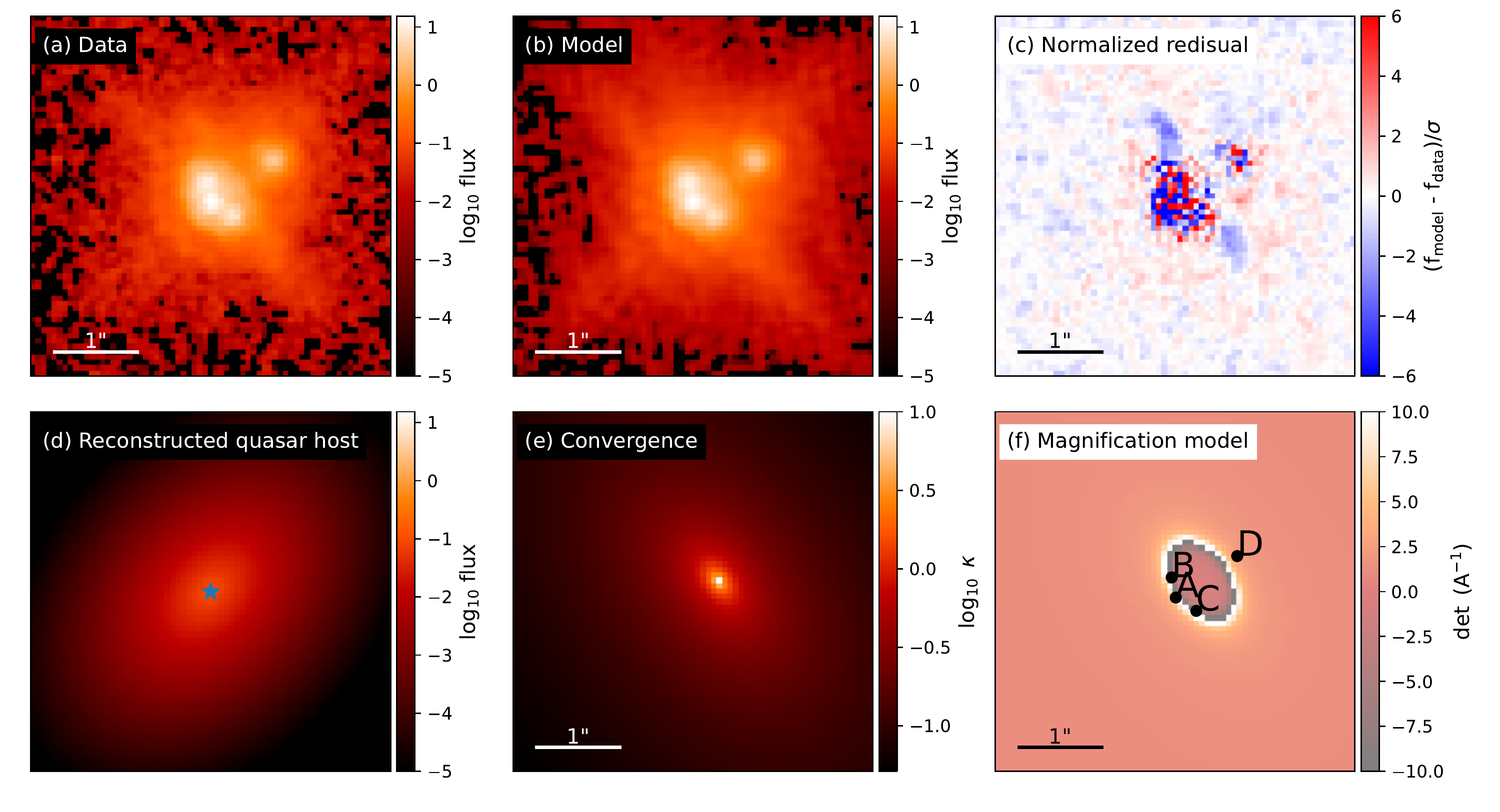}
\includegraphics[width=\linewidth]{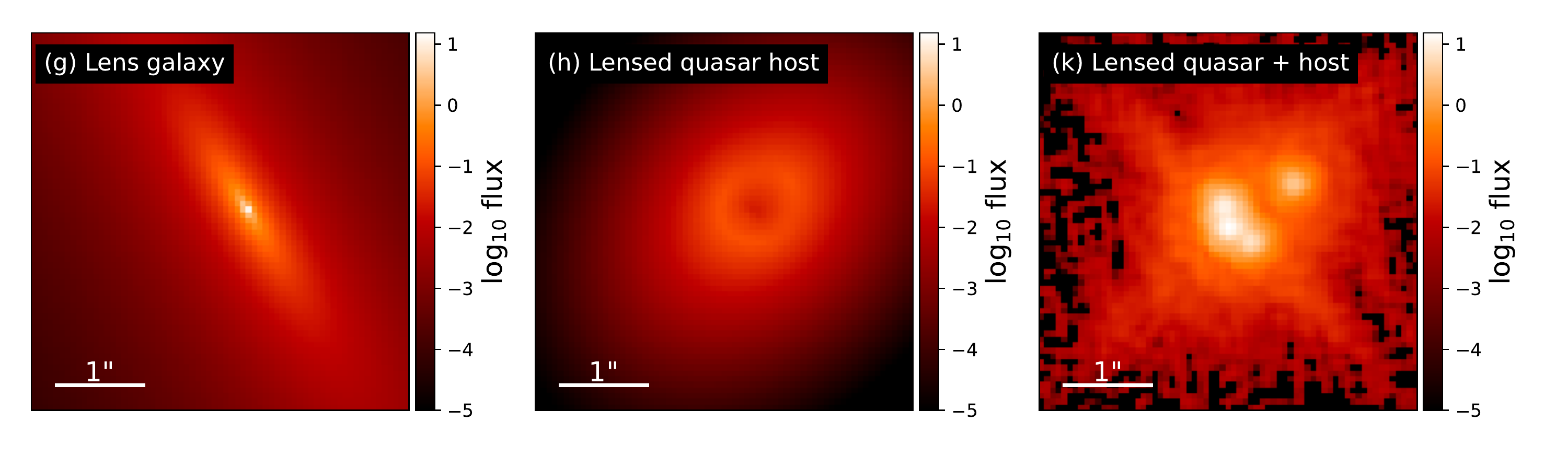}
\caption{Comparison of the (a) observed image with the (b) reconstructed image from lens modeling. Also shown are the (c) normalized fitting residual, the (d) reconstructed quasar position (marked by the star) and its host galaxy, a plot of the (e) unitless convergence and (f) magnification, as well as the (g) extended surface brightness of the foreground lens, the (h) lensed quasar host galaxy, and the (k) lensed quasar + quasar host. Model images shown in Panels (b) and (k) have been convolved with the PSF.}
\label{fig:lens}
\end{figure*}

Figure~\ref{fig:lens} (using the ePSF model) and Table~\ref{table:lens} (combining the results in all runs using different PSF models as in Table \ref{table:decomp}) summarize the lens modeling results. Our lens model achieved a satisfactory fit to the data and successfully reconstructed the observed image configuration, yielding a goodness-of-fit comparable to or slightly better than that obtained from image decomposition in all runs. The extended emissions surrounding G1 in Figure \ref{fig:decomp} were attributed to the lensed quasar host galaxy (Panel (h) in Figure \ref{fig:lens}) in the lens modeling. However, we note that the reconstruction of the unlensed quasar host galaxy is uncertain and differs significantly in different runs. This is likely due to the intrinsic faintness of the quasar host ($\sim23$ mag), as is typical at $z\sim2.5$, and its blending with ABCD and G1. For example, in some runs, the residual flux of the warped disk was treated as part of the lensed quasar host galaxy and biased the reconstruction. Nonetheless, the constraints on the mass profile of G1 and the magnification model remain robust since they are mainly derived from the positions of the bright point sources. 

\begin{figure}
\includegraphics[width=\linewidth]{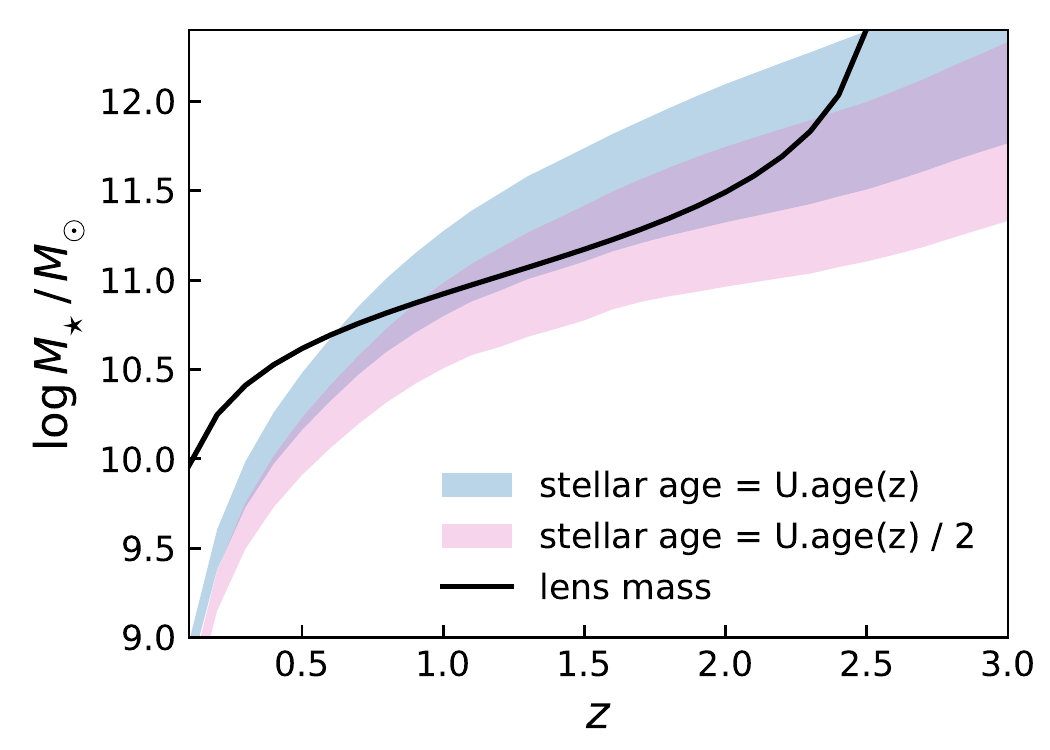}
\caption{Comparison of the stellar mass of the foreground lens estimated from lens modeling (black curve) and stellar population analysis (shaded regions) as a function of the assumed  redshift of the lens. The age of the main stellar population is fixed to either the age (blue) or half the age (pink) of the universe at a given redshift. The lower and upper edges of the shaded region correspond to $\ebvline=0.0$ and $\ebvline=1.0$ assumed in the stellar model, respectively.}
\label{fig:lens_mass}
\end{figure}

The main deflector of the system is the edge-on disk galaxy G1 with an Einstein radius of $\sim0.42''$. The high inclination of G1 makes it an effective lens since edge-on systems have higher projected mass density and lensing cross section \citep[e.g.,][]{Maller1997, Keeton1998}. As discussed in Section \ref{subsec:decomp}, its disk is warped which may result from the interaction with its satellites in the past. Although most of the foreground lenses identified thus far have been massive ellipticals, there is an increasing number of edge-on disk lenses or disturbed/interacting lenses being discovered, either through dedicated surveys or serendipitous observations \citep[e.g.,][]{Suyu2009, Sygnet2010, Treu2011}. Therefore, the detection of an irregular “host galaxy” can no longer be served as a smoking gun evidence for dual/multiple AGNs triggered in galaxy mergers, since the foreground lens could also have a disturbed morphology.

Assuming a redshift for the foreground lens G1, its mass within the Einstein radius, presumably dominated by stars in the central $\rm \sim kpc$ region of the galaxy, can be estimated as 
\begin{equation}
    M(\thetae) = \frac{c^2}{4G}\frac{D_s D_d}{D_{ds}} \thetae^2, 
\end{equation}
where $D_s$, $D_d$, $D_{ds}$ are the angular diameter distances of the source, the lens, and that between the lens and the source, respectively. Alternatively, the stellar mass of G1 can be estimated from the decomposed F160W magnitude ($\sim20.8$ mag) through a stellar population synthesis analysis. Adopting a \cite{Bc03} stellar population model, a \cite{Chabrier2003} initial mass function, and a \cite{Calzetti2000} extinction law ($\ebvline = 0.0 - 1.0$), the inferred stellar mass of G1 as a function of redshift, compared to that estimated from the lens modeling, is shown in Figure \ref{fig:lens_mass}. There is a broad redshift and parameter range over which the stellar masses derived from the two methods are consistent. This demonstrates that our lens modeling produced a lens galaxy with a reasonable mass-to-light ratio. Considering the rapid drop off of the stellar mass function beyond $\logm/\msun \sim 11.0$, the foreground lens might be located at $z\sim1.0$ with  $\logm/\msun \sim 10.8$.

The total magnification of this system predicted from the model is $\sim32$, and the predicted flux ratio (relative to A) of $\rm A:B:C:D$ is $1.00_{-0.00}^{+0.00}:0.53_{-0.01}^{+0.00}:0.46_{-0.00}^{+0.00}:0.17_{-0.00}^{+0.01}$. The observed flux ratio slightly differs from the model prediction for component B (which also exhibits slight spectral variations in comparison to A and C in \citealt{Ciurlo2023}), with the actual values being $1.00_{-0.00}^{+0.00} :0.69_{-0.02}^{+0.01} :0.45_{-0.00}^{+0.00} :0.23_{-0.01}^{+0.01}$. Notably, the continuum flux ratio deviates from the \ha~flux ratio reported in \cite{Ciurlo2023}, which is approximately $1.0:0.5:0.25$ for $\rm A:B:C$. There are several factors that could cause the flux ratio anomalies, including 1) substructures that are not considered in our lens model; 2) imperfect subtraction of multiple point sources from the underlying (disturbed) galaxies, which is known to be a challenge even for single quasars; 3) microlensing by stars in the foreground galaxy; and 4) impact of differential dust extinction. 

In the scenario where the foreground lens is a warped disk galaxy, the non-smooth distribution of baryonic components that causes perturbations in the lensing potential can be a major contribution to the flux ratio anomalies \cite[e.g.,][]{Hsueh2018}. There is also an enhanced likelihood of microlensing by stars and extinction by dust when multiple sources are located near the center of the lens galaxy. Microlensing effects could cause the slight spectral differences observed in \cite{Ciurlo2023} and the flux ratio anomalies between the continuum and the \ha~line \cite[e.g.,][]{Braibant2014}, since they affect the compact continuum emission region and the more extended broad-line region differently \citep[e.g.,][]{Keeton2006}. Differential dust extinction within the lens galaxy can result in varying degrees of extinction and flux reduction for individual lensed images. Multiband high-resolution imaging and the redshift of foreground lens are necessary to correct the dust extinction effect for more accurate lens modeling \cite[e.g.,][]{Suyu2009}.

\subsection{Multiple quasars triggered in galaxy mergers?}
While the lensing model provides a self-consistent explanation of the observations, the scenario of multiple AGNs triggered in galaxy mergers cannot be completely ruled out. The foreground lens could otherwise be the disturbed host galaxies of multiple distinct AGNs at $z\sim2.58$, with a stellar mass larger than $\sim10^{11}\,\msun$ (Figure \ref{fig:lens_mass}).

In the general merger picture, if multiple galaxies, each hosting a SMBH, are involved in a merger event, the formation of a multiple SMBH system is inevitable. However, simultaneous activation of all SMBHs as AGNs is rare \citep[e.g.,][]{Benitez2023, Foord2021}. Cosmological hydrodynamic simulations now reach a sufficient volume to allow a statistical assessment of the occurrence rate of dual/multiple AGNs. In the ASTRID simulation \citep[e.g.,][]{Hoffman2023, Chen2023_simu}, although $\sim3\%$ of massive BHs ($\mbh>10^7\,\msun$) at $z\sim2.5$ are in a triple SMBH system with projected separation $r_p < 120$~kpc, the AGN activity of the third (faintest) BH is usually weak with $\lbol \lesssim 10^{44}\,\ergs$, especially at close separations ($r_p < 40$~kpc) where the third BH can be strongly deactivated through gas stripping and tidal disruptions. Overall, only $\sim0.1\%$ of the triple systems at cosmic noon in ASTRID have $\lbol\gtrsim10^{45}\, \ergs$ for each member at $r_p < 40$~kpc. Similarly, in the Horizon-AGN simulation, quadruple systems with each AGN brighter than $\lbol > 10^{45}\,\ergs$ (like J1608) and separated by $r_p<30$ kpc only constitute $\sim0.001\%$ of the entire AGN population at $z\sim2.5$ \citep{Volonteri2022}. The realistic probability of observing a luminous quadruple quasar system with $r_p \lesssim 7$~kpc ($\sim0.9''$) would be even lower. Another issue with the merging scenario is that none of the triple ABC are centered on the most luminous galaxy G1. This might be expected for high redshift clumpy galaxies \citep[e.g.,][]{Bournaud2012} but overall the morphology of G1 is not that irregular. Finally, the classic cusp configuration of the four point sources strongly favors the lensing scenario, as it is highly unlikely for a multiple quasar system to be aligned in this particular configuration. 

Therefore, the probability of J1608 being a bona fide quadruple quasar system is extremely low, if not impossible.

\section{Conclusions}
We present HST/WFC3 deep IR imaging in the F160W band of SDSS J1608+2716, which provides new insights into the nature of this unique system. Previous Keck AO-assisted IFU spectroscopy have revealed three spatially distinct point-source components separated by $0.25''$ (corresponding to $\sim2$ kpc at $z = 2.575$), making it a promising candidate for a close-separation triple quasar at cosmic noon. Our deep HST imaging has uncovered a fourth point-like component at a distance of $\sim0.9''$ from the triple, and a disk galaxy  located in between the four point sources that was not detected in the shallow Keck imaging. The disk galaxy displays a U-type warped morphology which is likely being tidally stripped by its satellite galaxies.
The entire system exhibits a characteristic cusp structure that resembles the image configuration of a quadruply lensed quasar, and the source positions can be successfully reproduced by a singular isothermal ellipsoid lens model. These compelling findings indicate that J1608 is a single quasar being lensed into four images.  

Our result demonstrates the challenge of finding dual/multiple AGNs with close separations at high redshifts, and have important implications for ongoing efforts. It can be seen that even with superb resolution achieved by AO-assisted Keck observations, its small FOV and/or low sensitivity may miss some of the fainter/additional lensed images and/or the foreground lens. Moreover, the detection of an irregular ``host galaxy'' does not guarantee that the system is a dual/multiple AGN, because the foreground lens could also have a disturbed morphology. This underscores the importance of deep, high-resolution IR imaging and the caveats with limited data to robustly confirm the dual/multiple AGN nature at high redshift.

In the imminent future, the Nancy Grace Roman Space Telescope is anticipated to deliver deep (reaching $J\sim26-27$ mag) and high-resolution ($\sim0.11''$) multi-band imaging ($\rm 0.5-2.3\,\mu m$) for millions of galaxies over $\sim2000\rm\ deg^{-2}$, supplemented by the low-resolution grism spectroscopy with its High Latitude Wide Area Survey \citep{Wang2022}. This survey will revolutionize the discovery and identification of subarcsecond dual/lensed systems at high redshifts, enabled by simultaneously searching for multiple point sources in close proximity, eliminating star-quasar superpositions via photometric color and spectroscopic information, detecting the putative foreground lens and measuring its redshift, and depicting the morphology of the faint quasar host galaxies. Complemented by the unparalleled observing capability of JWST to characterize the detailed properties of individual targets, we are now entering a new era to build statistically significant samples of dual AGNs to quantify their abundance as functions of separation, luminosity, redshift, and host galaxy properties \citep[e.g.,][]{Shen2023}. This will, in turn, provide essential observational constraints on the physical mechanisms driving AGN fueling and SMBH growth in the general framework of galaxy evolution.

\begin{acknowledgments}
We thank A. Pagul and A. Vick for help with our HST observation. We thank the anonymous referee for giving constructive comments which helped improve the quality of the paper. This work is supported by NSF grant AST-2108162. YS acknowledges partial support from NSF grant AST-2009947. Support for Program number HST-GO-17287 was provided by NASA through grants from the Space Telescope Science Institute, which is operated by the Association of Universities for Research in Astronomy, Incorporated, under NASA contract NAS5-26555. This work was supported by JSPS KAKENHI Grant Numbers JP22H01260, JP20H05856, JP20H00181.
Based on observations made with the NASA/ESA Hubble Space Telescope, obtained from the Data Archive at the Space Telescope Science Institute, which is operated by the Association of Universities for Research in Astronomy, Inc., under NASA contract NAS 5-26555. These observations are associated with program GO-16892, GO-17287, GO-17269 (PI: X. Liu). The HST data used in this paper can be found in MAST\dataset[DOI]{http://dx.doi.org/10.17909/b92g-sp96}.
\end{acknowledgments}

\software{\texttt{Astropy} \citep{astropy}, \texttt{DrizzlePac} \citep{Hoffmann2021}, \texttt{lenstronomy} \citep{Birrer2018, Birrer2021},
\texttt{photutils} \citep{Bradley2022}} 

\bibliography{sample631.bbl}

\end{document}